# MODEL OF FORMATION OF MONODISPERSED COLLOIDS

## Jongsoon Park, Vladimir Privman and Egon Matijević[*]

Center for Advanced Materials Processing, Clarkson University

Potsdam, New York 13699–5820, USA

**ABSTRACT**

Ample experimental evidence has been accumulated demonstrating that the formation of monodispersed colloids proceeds through a more complex mechanism, than the generally excepted diffusional "burst nucleation" process. Instead, the synthesis of narrow-size-distribution colloidal dispersions involves several stages, i.e., nuclei produced in a supersaturated solution, grow to nanosize subunits, which then aggregate to form much larger uniform secondary particles. To explain the size selection in such a series of processes, a kinetic model was developed which combined two growth/aggregation stages.

This earlier study has shown the burst-nucleation growth of the primary particles to depend strongly on the value of the effective surface tension entering the surface term in the free energy of the subcritical embryos. The aim of the present work has been to identify an appropriate control parameter in the process of secondary particle aggregation. Modifications of the aggregation rates to account for singlet size and aggregate diffusivity produced only small changes. However, introduction of a "bottle-neck" factor in the dimer formation rate had a profound effect on the final size distribution and suggested a possible size-control mechanism.

---

[*]corresponding author

# 1. INTRODUCTION

This work revisits our recently developed model,[1] of growth of monodispersed colloids by precipitation from homogeneous solutions,[2-7] and reports new theoretical investigations aimed at improving the agreement of the model results with experiment. A large number of dispersions of uniform colloid particles of various chemical composition and shape, ranging in size from fraction of a micron to few microns, have been described. Modeling of their formation relies on experimental identification of the solute species that are involved in the various stages of the colloid particle formation and on observation of growth stages leading from the initial nucleation to the final particles. Early theoretical modeling was based on the mechanism suggested by Reiss,[8] and by LaMer and Dinegar,[9-10] which assumes a short nucleation burst, followed by diffusional growth of the resulting nuclei to form identical fine particles. Utilization of related ideas to initiate the field of study of defect association in semiconductors was developed by Reiss and coworkers.[11-12]

This "burst-nucleation" mechanism works well for particles up to several tens on nanometers in diameter, though the resulting peaked size distribution is not very narrow; the width is comparable to the average size. For larger particles, of sizes from order 100 nm up, there has been mounting experimental evidence that the burst-nucleation/diffusional growth mechanism alone is inadequate. Specifically, it has been found that many spherical particles precipitated from solution showed polycrystalline X-ray characteristics, such as ZnS,[13] CdS,[14], $Fe_2O_3$,[15] etc. These particles are not single crystals. Rather, it has been confirmed by several techniques: small angle light scattering, electron microscopy, X-ray diffraction, that most monodispersed colloids consist of small crystalline subunits.[13-23] Furthermore, it has been observed,[1,7,19] that the crystalline subunits in the final particles were of the same size as the diameter of the precursor-subunit "singlets" of sizes of order 10 nm, formed in solution, thus suggesting an aggregation-of-subunits mechanism. This two-stage growth process is shown schematically in Figure 1. Figures 2 and 3 illustrate the gold particles which were used as the experimental system,[1] to test the theoretical model. The composite structure has also been identified in non-spherical particles, and it has been recognized that different morphologies of the final particles must be related to the nature of the precursor singlets.[15,21–23]

The experimental evidence poses two theoretical challenges. First, the morphology and shape selection of particles formed by interplay of nucleation and aggregation processes must be explained. Second, the size-selection mechanism, i.e., the kinetics of generation of narrow particle size distribution, must be identified. Several theoretical approaches utilizing thermodynamic and dynamical growth mechanisms,[1,4,5,21,24–35] have been described in the literature. Models of aggregation of subunits can be developed that yield a peaked and even sharpening with time particle size distribution.[24–37] However, none of the earlier attempts could fit quantitatively a broad range of experimental findings. Here we review and further develop our new approach,[1] that explains the *size selection*, by coupling the dynamical equations for the processes of secondary particle growth by aggregation of subunits and of formation (nucleation) of these subunits.

The main new finding of our work,[1] crucial to obtaining narrow size distribution, has been that the growth of the final, secondary particles by aggregation of singlets, must be coupled, via the rate of formation of these primary particles (singlets), to the time-dependence of the process of their nucleation and growth. We took the simplest possible models of both processes (primary and secondary particle formation), because this simplifies numerical calculations and avoids introduction of unknown microscopic pa-rameters; as a result we only fit *one* parameter, the effective surface tension, and even that parameter turns out to be close to the experimental bulk value. In Section 2, primary particle (singlet) formation by burst nucleation is described. The surface tension parameter enters the modeling of this process. Our aim in this work has been to identify a parameter which, respectively, has significant effect on the secondary particle growth by aggregation. In Section 3, a kinetic model of secondary particle formation by singlet-capture dominated growth is considered. The modifications of the rate expressions yielding the control parameter related to the "bottle-neck" suppression of the formation of dimers (doublets) from singlets, is presented in Section 4 which reports results of numerical simulations and comparison with an experiment.

## 2. BURST NUCLEATION OF THE PRIMARY PARTICLES

In modeling the formation of the primary particles, the burst-nucleation approach,[8–10] can be used. Here we briefly summarize the notation and parameter values involved, following our earlier work.[1] This approach requires modeling of the free energy $\Delta G$ of the growing embryos. We use the standard volume plus surface energy expression. We will refer to the species (atoms, ions, molecules) which serve as monomers for the primary-particle nucleation as *solutes*. For a given concentration $c(t)$ of the solutes, larger than their equilibrium saturation concentration $c_0$ and approaching $c_0$ for large times $t$, the rate of formation of critical nuclei, per unit volume, can be written[38,39] as

$$\rho(t) = 4\pi a n_{cns}^{1/3} \mathcal{D} c^2 e^{-\Delta G_{cns}/kT}, \qquad (1)$$

which is based on the diffusional capture of solutes, whose effective radius is denoted by $a$, diffusion constant by $\mathcal{D}$, and $n$ is the number of solutes in an embryo, with the subscript *cns* referring to values calculated at the critical nucleus size. Let us summarize the assumptions that yield this relation.

For embryos of size $n < n_{cns}$, the solutes can be captured and emitted fast enough so that the size distribution is given by the equilibrium form. Thus, the factor $ce^{-\Delta G_{cns}/kT}$ in Eq. 1 follows from the expectation that embryo sizes up to $n_{cns}$ are thermodynamically distributed, according the Boltzmann factor. For sizes larger than $n_{cns}$, the dynamics is assumed to be fully irreversible, corresponding to unbounded growth by capture of solutes. It is assumed that the $n$-solute embryo (cluster) has radius $an^{1/3}$, which defines the effective solute radius $a$, including the volume filling-fraction correction. The factor $4\pi a n_{cns}^{1/3} \mathcal{D} c$ in Eq. 1 corresponds to the Smoluchowski diffusive capture rate of solutes at the critical nucleus radius.[40]

In order to test the model, we used the experimental data on the growth of dispersions of submicron spherical gold particles,[1,7] which were produced by the reduction of chloroauric acid ($HAuCl_4$) with ascorbic acid. The effective radius of the gold atom, $a = 1.59 \cdot 10^{-10}$ m, was obtained by dividing the actual radius[41,42] of the gold atom, $1.44 \cdot 10^{-10}$ m, by the cubic root of the volume filling fraction, 0.74, of the crystalline structure of gold.[43] The solute diffusion coefficient was estimated by using the Stokes-Einstein formula with the actual radius of the gold atom, and the result was $\mathcal{D} = 1.5 \cdot 10^{-9}$ m$^2$sec$^{-1}$.

For the free energy of the $n$-solute embryos, the following expression can be used,

$$G(n) = -nkT \ln(c/c_0) + 4\pi a^2 n^{2/3} \sigma, \qquad (2)$$

which involves the bulk term, proportional to $n$, and the surface term. This expression is only meaningful for large $n$. Therefore, we put $\Delta G = G(n)$, ignoring the difference $G(1)$. Both $n_{cns}$ and $\Delta G_{cns}$ are thus explicit functions of $c(t)$,

$$n_{cns} = \left[\frac{8\pi a^2 \sigma}{3kT \ln(c/c_0)}\right]^3, \tag{3}$$

$$\Delta G_{cns} = \frac{256\pi^3 a^6 \sigma^3}{27(kT)^2 \left[\ln(c/c_0)\right]^2}, \tag{4}$$

where the critical value $n_{cns}$ was calculated from $\partial G/\partial n = 0$.

The surface term in Eq. 2 corresponds to the assumption of spherical nuclei of radius $an^{1/3}$ and introduces the effective surface tension $\sigma$. The value of the effective surface tension of nanosize gold embryos in solution profoundly affects the numerical results. Unfortunately, even the bulk-gold value is not well known. It is of the order of 0.58 to $1.02\,\text{Nm}^{-1}$, where the lower value is more probable.[44] We use $\sigma$ as an adjustable parameter in our model.

The bulk term in Eq. 2 is purely entropic, assuming noninteracting-solute (dilute) solution. The saturation concentration of gold in solution, $c_0$, is not well known. Using $2 \cdot 10^{-12}\,\text{mol}\,\text{dm}^{-3}$,[45] yields $c_0 \simeq 1 \cdot 10^{15}\,\text{m}^{-3}$. The temperature is given as $kT = 4.04 \cdot 10^{-21}\,\text{J}$.

The formation and subsequent growth of the critical nuclei cause the decrease in the concentration of solutes, $c(t)$. In the burst nucleation model,[9,10] it is assumed that fewer nuclei are produced at later times, while the existing nuclei grow at the expense of the remaining solutes and sub-critical embryos. The size distribution then develops a peak but it is not as sharp as observed for the larger secondary particles. Furthermore, as mentioned in the introduction, there is ample evidence that the secondary particles are compound.

In the present approach, we assume for simplicity that the primary particles are captured fast enough by the growing secondary particles so that the effect of their aging on the concentration of solutes can be ignored. Furthermore, the radius of the captured primary particles will be assumed close to the critical radius. We discuss the implications of these assumptions in the next section. Thus, we write

$$\frac{dc}{dt} = -n_{cns}\rho(t),  \qquad (5)$$

which means that the concentration of solutes is "lost" solely due to the irreversible formation of the critical-size nuclei. Collecting all the above expressions, one gets the following equations for $c(t)$ and $\rho(t)$,

$$\frac{dc}{dt} = -\frac{16384\pi^5 a^9 \sigma^4 \mathcal{D} c^2}{81(kT)^4 \left[\ln\left(c/c_0\right)\right]^4} \exp\left\{-\frac{256\pi^3 a^6 \sigma^3}{27(kT)^3 \left[\ln\left(c/c_0\right)\right]^2}\right\}, \qquad (6)$$

$$\rho(t) = \frac{32\pi^2 a^3 \sigma \mathcal{D} c^2}{3kT \ln\left(c/c_0\right)} \exp\left\{-\frac{256\pi^3 a^6 \sigma^3}{27(kT)^3 \left[\ln\left(c/c_0\right)\right]^2}\right\}, \qquad (7)$$

where the initial concentration of the gold solution used in the preparation of the dispersion was $c(0) = 6.0 \cdot 10^{25}$ m$^{-3}$.

## 3. FORMATION OF THE SECONDARY PARTICLES BY AGGREGATION

The growth of the secondary particles is facilitated by the appropriate chemical conditions in the system: the ionic strength and/or pH must assume values such that the surface potential approaches the isoelectric point, resulting in reduction of electrostatic barriers, thus promoting fast irreversible particle aggregation. Formation of the final (secondary) particles, which can be of narrow size distribution, is clearly a diffusion-controlled process.[1-7]

We describe the process by a master equation for the distribution of growing particles by their size. Here it is assumed that the particles are spherical, with the density close to that of the bulk material. Experimentally, the growing particles rapidly restructure to assume the final shape and density, so they are not fractal even though the transport of the constituent units is diffusional. The modeling of this restructuring is an interesting unsolved problem on its own, but, as long as the restructuring is fast, its mechanism plays no role in the master equation.

The cluster size will be defined by how many primary particles (singlets) were aggregated into each secondary particle, denoted by $s = 1, 2, \ldots$. The growing particles can adsorb or emit singlets and multiplets. The master equation is then quite standard to set up.[46] In the present case, the singlets have diffusion constants larger than aggregates so that their capture dominates the growth process. This "singlet dominance" has been traditionally assumed in the literature,[26] mostly because it decreases the number of parameters in the model and simplifies numerical calculations. The master equation, assuming no detachment, is

$$\frac{dN_s}{dt} = K_{s-1} N_1 N_{s-1} - K_s N_1 N_s \qquad (s \geq 3). \tag{8}$$

Here $N_s(t)$ is the time-dependent number density (per unit volume) of the secondary particles consisting of $s$ primary particles; $K_s$ is the attachment rate constant.

Here we assume the Smoluchowski rate expression,[26,40]

$$K_s = 4\pi (R_1 + R_s)(D_1 + D_s), \tag{9}$$

where $R_s$ is the radius of the $s$-singlet particle, given by

$$R_s = 1.2\,rs^{1/3}\,. \tag{10}$$

and $D_s$ is the diffusion coefficient of the secondary particle of size $s$,

$$D_s = \frac{D_1}{s^{1/3}}\,. \tag{11}$$

The average radius of the primary particles, $r$, and their diffusion coefficient, $D_1$, are experimentally available, the latter via the Stokes-Einstein relation: $r = 2.10 \cdot 10^{-8}$ m, $D_1 = 1.03 \cdot 10^{-11}$ m$^2$sec$^{-1}$, for our selected experimental system. The constant 1.2 in Eq. 10 was calculated as

$$(0.58)^{-1/3} \simeq 1.2\,, \tag{12}$$

where 0.58 is the typical filling factor of the random loose packing of spheres.[43]

The concentrations of monomers (singlets) and dimers are not covered by Eq. 8. For dimers, we add the factor $f = 1/2$ to the monomer-monomer coagulation term owing to the double counting,

$$\frac{dN_2}{dt} = fK_1 N_1^2 - K_2 N_1 N_2\,. \tag{13}$$

The notation with variable $f$ will be explained in Section 4. The equation for the monomer concentration can be obtained using the conservation of matter and the rate of the primary particle formation, $\rho(t)$,

$$\frac{dN_1(t)}{dt} = \rho(t) - \sum_{j=2}^{\infty} j\frac{dN_j(t)}{dt}\,, \tag{14}$$

with the initial values $N_s(0) = 0$ for all $s = 1, 2, 3, \ldots$.

This latter assumption, that the secondary process is "fed" by the primary one, is crucial to obtaining narrow size distributions. The growth of the secondary particles must be coupled, via the rate of generation of singlets (primary particles), to the time-dependence of the process of formation of the latter.

In the original work,[1] we assumed one further simplification: we took the Smoluchowski rate corresponding to pointlike singlets and stationary aggregates, i.e.,

$$K_s = 4\pi R_s D_1, \qquad (15)$$

and we used Eq. 8 down to $s = 2$ because the rates for small $s$ are anyway quite uncertain. The latter approximation provides a smoother $s$-dependence at $s = 1$.

The choice Eq. 15 also makes more apparent the following important property. Let us point out that, actually for both choices of the rates, the radii involved are proportional to the singlet radius $r$. The diffusion constants are inversely proportional to $r$, according to the Stokes-Einstein relation. Therefore, the fact that singlet sizes are actually relatively widely distributed, plays little role is the rate constants for capture of different-size singlets. To a good approximation, this distribution in sizes can at most affect the value of the porosity-related constant 1.2 in Eq. 10. We therefore used the experimentally observed average singlet radius, given earlier.

A more important approximation was stated in Section 2: in the solute matter balance, we assumed that the critical nuclei do not grow much before they are captured by the secondary particles. While this simplifies the model, matter conservation can be violated, so our calculated distribution $N_s(t)$ should be regarded as relative. It must be normalized for the total amount of matter. The latter is usually easily done because the amount of the solute material produced in the initial "induction" step is known.

# 4. RESULTS, DISCUSSION, AND COMPARISON WITH EXPERIMENT

In Figures 4 through 6, the results of numerical simulations with the new aggregation rate Eq. 8, see lower panels, are compared with our previous work, upper panels, for three different values of $\sigma$. The final size distribution after the experiment duration time of 10,sec, depends very sensitively on the choice of the surface tension parameter $\sigma$. Actually, we scanned various values of $\sigma$ and found the one that corresponds to the largest average particle size, $\sigma = 0.57\,\text{Nm}^{-1}$, see Figure 5, which is surprisingly close to the bulk value of the surface tension, cited in Section 2.

For smaller values of $\sigma$, illustrated in Figure 4, the growth process reaches saturation at small particle sizes. For larger $\sigma$ values, illustrated in Figure 6, the growth process results in a broad distribution and is incomplete, on the time scales of relevance to the experiment. While the sharpness of the distribution and the estimate of $\sigma$ are consistent with the measured values,[1] the particle size is smaller than that experimentally observed, so the consistency with the experiment can be described as semiquantitative.

Our modification of the rate expressions, described in Section 3, did not affect the resulting particle sizes significantly. Let us describe now several observations that do follow from the numerical results with this modification, which mostly increases the small-$s$ rates. The new rates do yield somewhat larger particle sizes, and the particle size distribution is somewhat sharper. Furthermore, with the new rates we see clearly that the value of $N_1(t)$ is not a smooth continuation of the $s > 1$ size-distribution values. This will have implications for setting up continuous (in $s$) modeling equations which may be needed to allow future more extensive numerical studies with additional rates, notably, with detachment and cluster-cluster aggregation included.

Let us point out that the secondary particle growth is experimentally the most controlled and the slowest process in the multistage synthesis summarized in Figure 1. Therefore, it is natural to search for parameters in this stage of the process that affect significantly the resulting size distribution. Note that we already know one "sensitive" parameter, $\sigma$, for the primary-particle nucleation. In search of one for the secondary particle growth, we have tried to modify the singlet-singlet aggregation rate by varying the parameter $f$ in Eq. 13.

Actually, if the particle surfaces are slightly charged, assuming constant and small surface-potential approximation[47] suggests that both the electrostatic and dispersion interaction energies between particles of radii $R_s$ and $R_1$, as functions of $s$, are approximately proportional to $1/(1 + s^{-1/3})$. Therefore the resulting energy barrier will be smaller by the factor of $1/2$ for $s = 1$ as compared to large $s$. The expectation in this case is then $f > 1/2$, i.e., the *relative* rate of singlet-singlet aggregation will be suppressed *less* than that for singlet-cluster aggregation. However, our numerical results, exemplified in Figure 7, suggest that larger particle sizes, improving consistency with the experiment, are actually obtained for $f < 1/2$. The reason is probably kinetic, but we cannot offer any quantitative model at this time.

In summary, we modeled synthesis of submicron size polycrystalline colloid particles with the size distribution that is narrowly peaked at an average value corresponding to a large number of primary particles in a final secondary particle. For the experimental gold-sol system, the model has worked reasonably well: the average size, the width of the distribution, the time scale of the process, and the fitted effective surface tension were all semiquantitatively consistent with the measured or expected values. Our main conclusion has been that multistage growth models can yield size-selection as a kinetic phenomenon, which has been observed in a large number of experimental systems. We have also identified that the secondary particle growth is particularly sensitive to the singlet-singlet aggregation rate.

**Figure 1:** The two-stage growth process. Uniform particle formation in colloid synthesis usually proceeds as follows. Initially, solutes are formed to yield a supersaturated solution, leading to nucleation. The formed nuclei may further grow by diffusive mechanism. The resulting primary particles (singlets) in turn aggregate to form secondary particles. This latter process is sometimes facilitated by changes in the chemical conditions in the system: the ionic strength may increase, or the pH may change, causing the surface potential to approach the isoelectric point. Formation of the final (secondary) particles, which can be of narrow size distribution, is also a diffusion-controlled process.

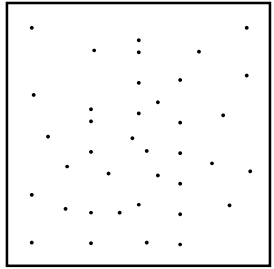

Supersaturated Solution

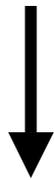

**Nucleation and Growth**

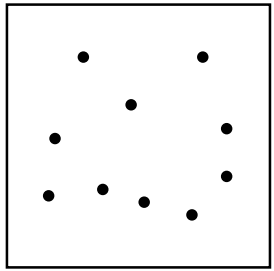

Primary Particles
(Monomers, Singlets)

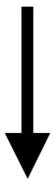

**Aggregation**

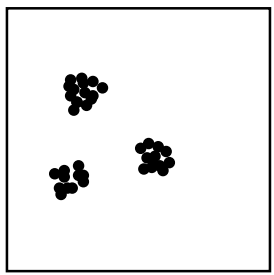

Secondary Particles
(Clusters, Aggregates)

**Figure 2:** Scanning electron micrographs, with increasing magnification from (A) to (B), of the final (secondary) gold particles obtained by rapidly adding $100\,\mathrm{cm^3}$ of an aqueous solution of $\mathrm{HAuCl_4}$ ($0.5\,\mathrm{mol\,dm^{-3}}$) into $400\,\mathrm{cm^3}$ of an aqueous solution of ascorbic acid ($0.5\,\mathrm{mol\,dm^{-3}}$), at room temperature.

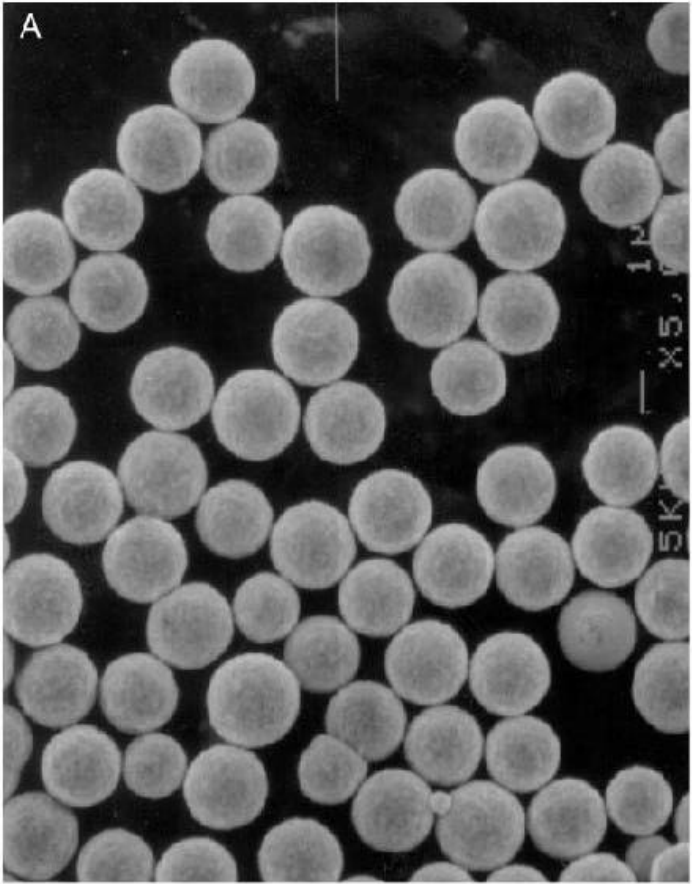
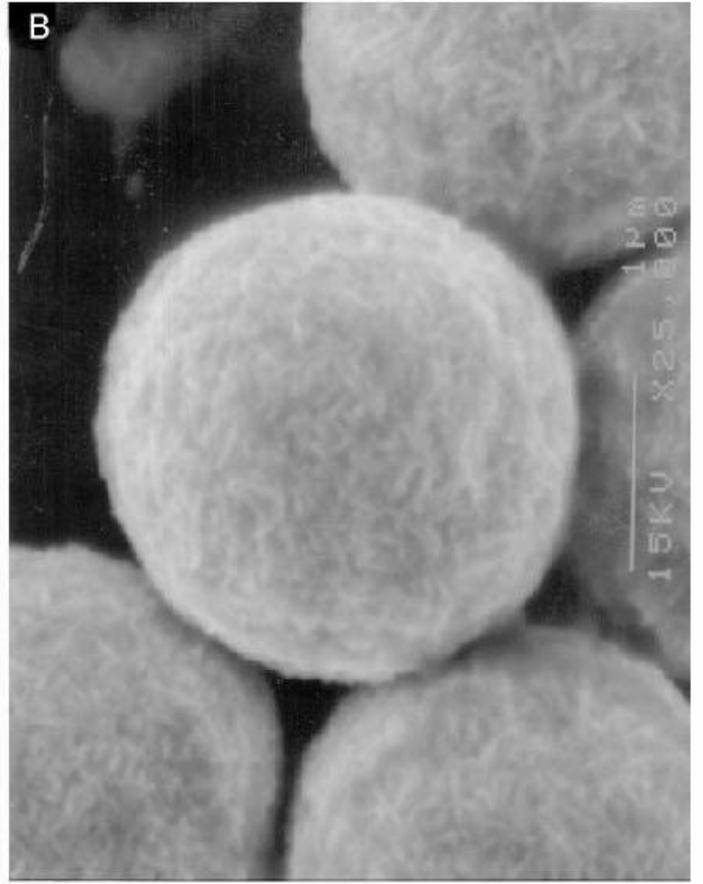

**Figure 3:** (A) Field emission microscopy image of gold particles shown in Figure 2. (B) Enlarged image of the darkened area in (A), which clearly reveals the presence of the subunits.

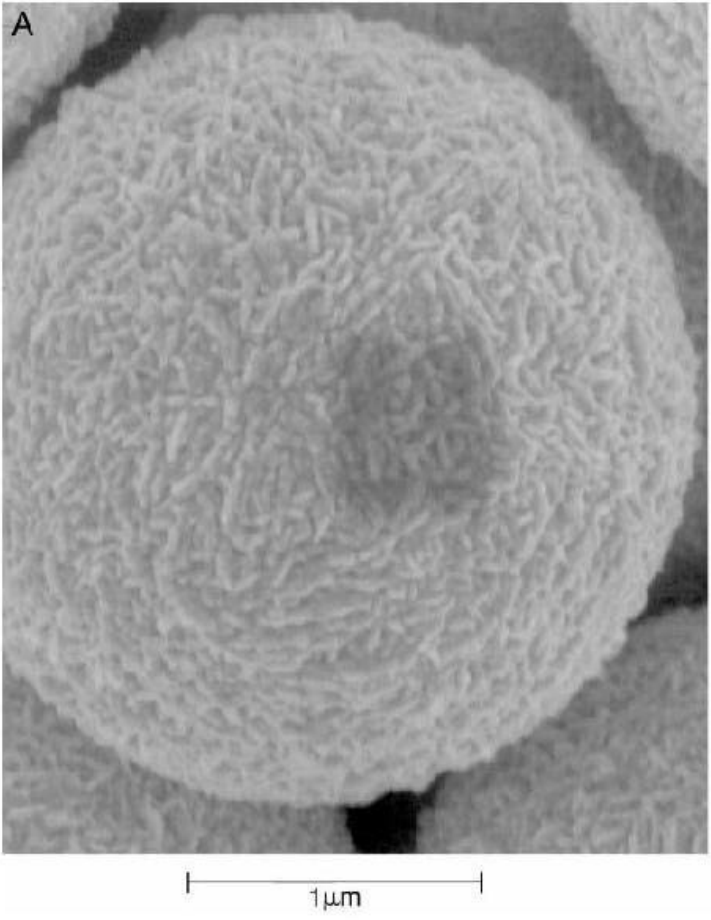 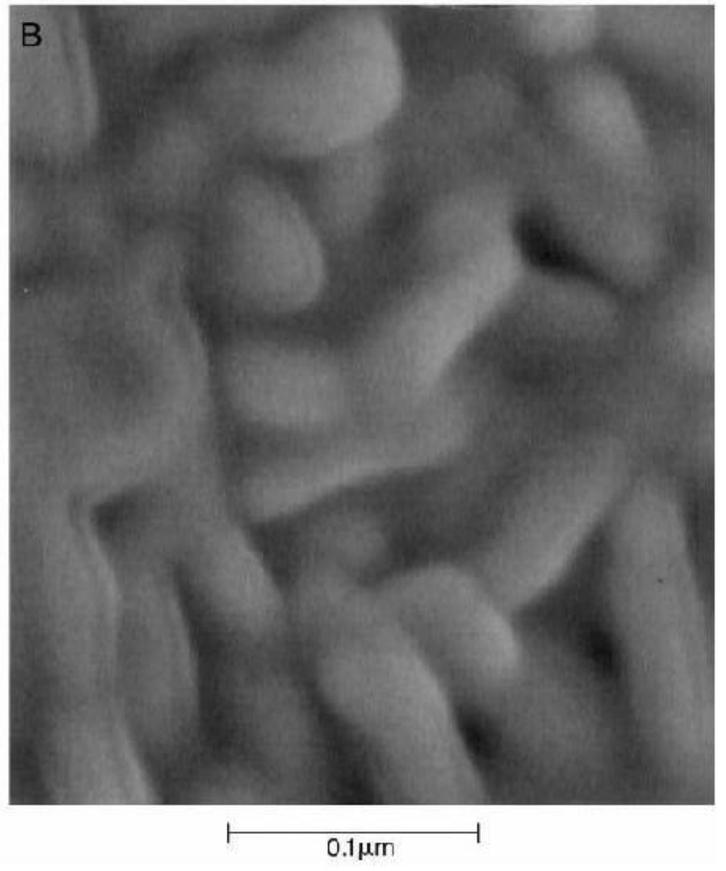

**Figure 4:** Distribution of the secondary particles by their sizes, calculated for times $t = 0.01, 0.1, 1, 10$ sec, using $\sigma = 0.51\,\text{N/m}$. The new rate expressions, see Eqs. [8,9,13], were used for the lower panel. The rates from Ref. (1) were used for the upper panel; see Section 3.

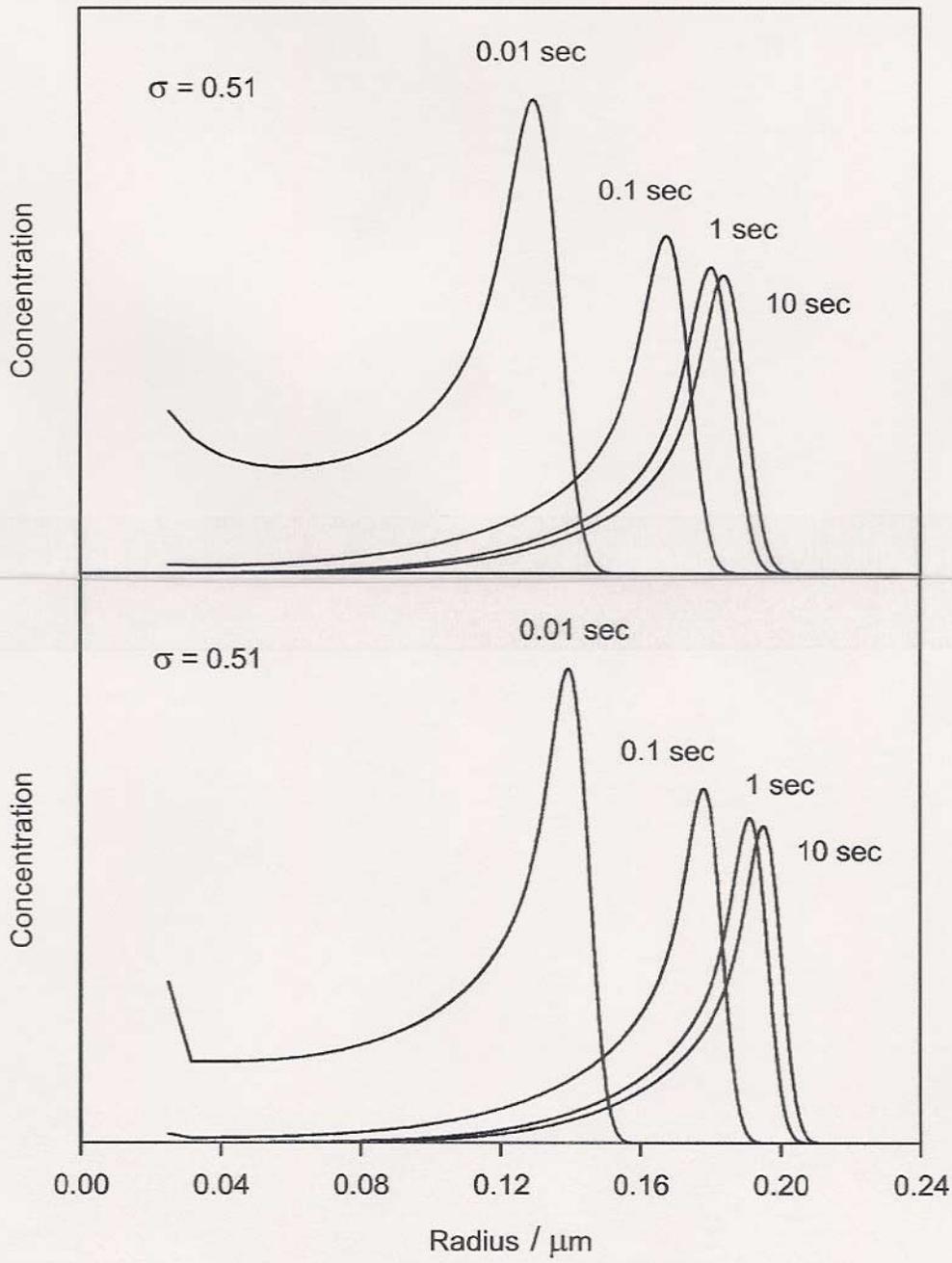

**Figure 5:** The same plot as in Figure 4, using $\sigma = 0.57\,\text{N/m}$.

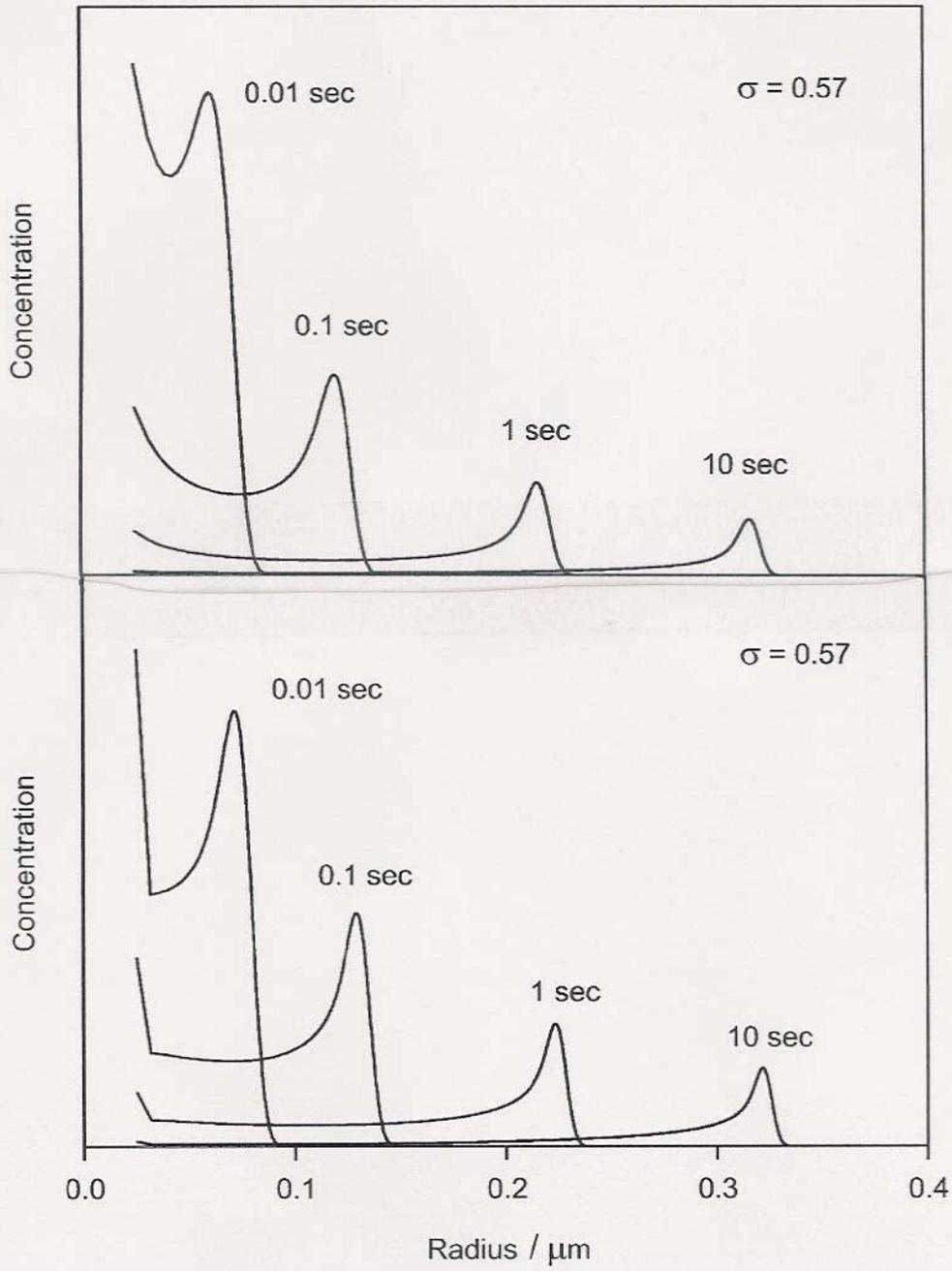

**Figure 6:** The same plot as in Figure 4, using $\sigma = 0.63\,\text{N/m}$, for times $t = 0.1,\ 1,\ 10$ sec.

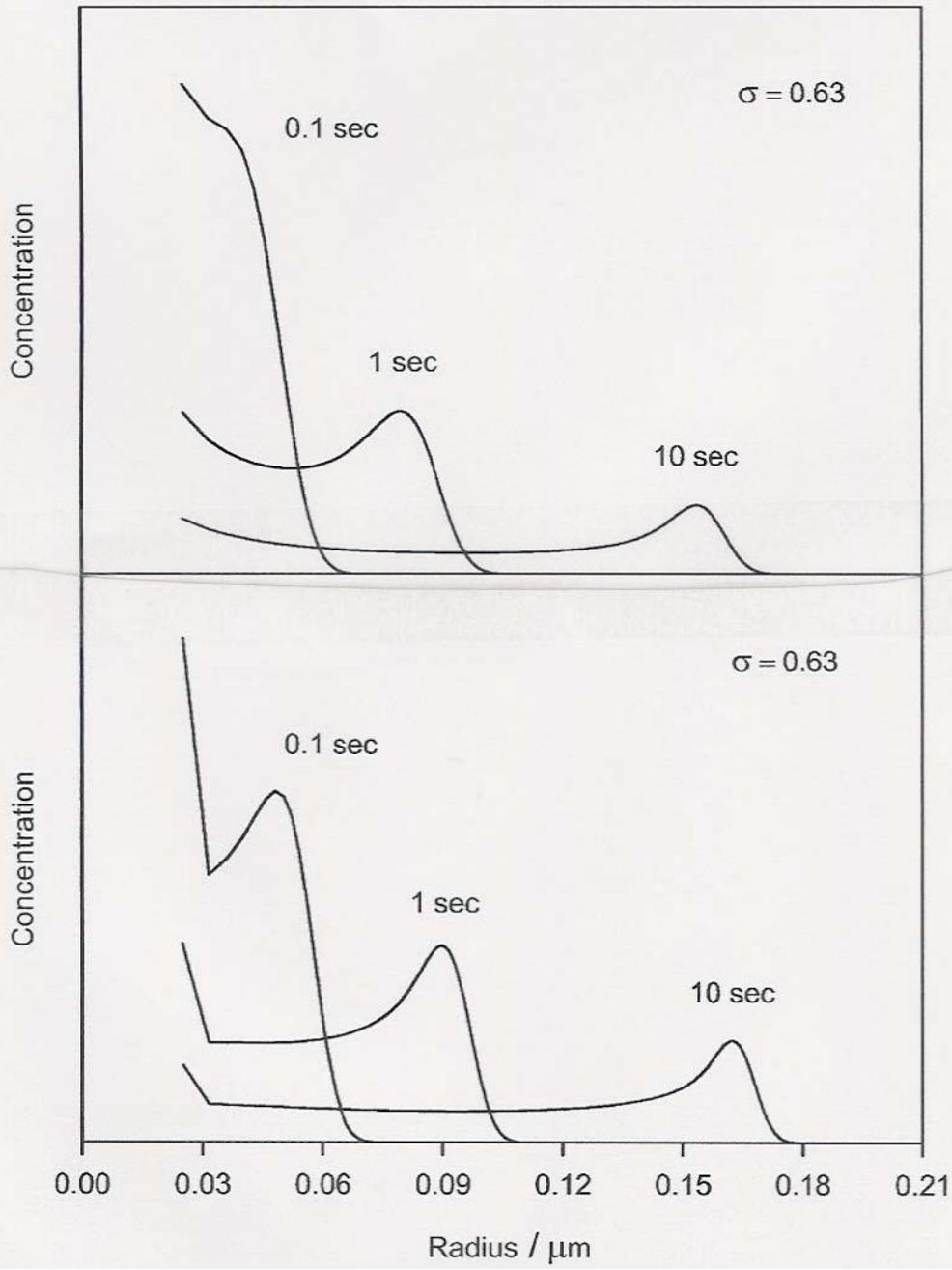

**Figure 7:** The secondary particle size distribution at time $t = 10\,\text{sec}$, for $\sigma = 0.57\,\text{N/m}$, for the choices $f = 0.1, 0.3, 0.5, 0.7, 0.9$, of the singlet-singlet rate modifying parameter $f$; see Section 4.

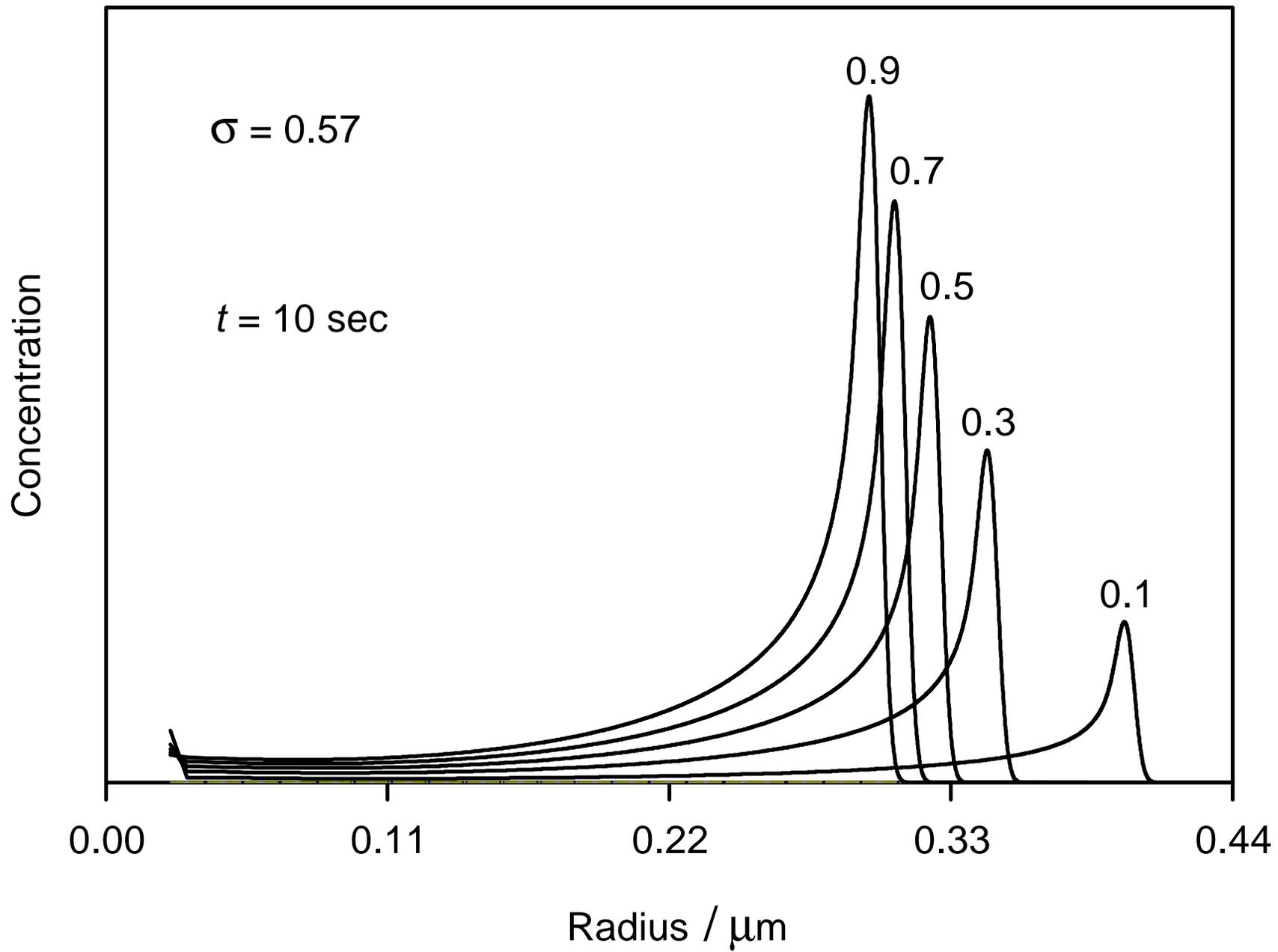